\documentclass[12pt,a4paper]{article}

\usepackage{ifthen} 
\newboolean{pdflatex}
\setboolean{pdflatex}{true} 

\newboolean{articletitles}
\setboolean{articletitles}{true} 

\newboolean{uprightparticles}
\setboolean{uprightparticles}{false} 

\usepackage{lineno}  
\usepackage{xspace} 

\usepackage{graphicx}  
\usepackage{color}
\usepackage{colortbl}
\graphicspath{{./figs/}} 

\usepackage{amsmath} 
\usepackage{amssymb}
\usepackage{amsfonts}
\usepackage{upgreek} 

\newcommand*\patchAmsMathEnvironmentForLineno[1]{%
\expandafter\let\csname old#1\expandafter\endcsname\csname #1\endcsname
\expandafter\let\csname oldend#1\expandafter\endcsname\csname
end#1\endcsname
 \renewenvironment{#1}%
   {\linenomath\csname old#1\endcsname}%
   {\csname oldend#1\endcsname\endlinenomath}%
}
\newcommand*\patchBothAmsMathEnvironmentsForLineno[1]{%
  \patchAmsMathEnvironmentForLineno{#1}%
  \patchAmsMathEnvironmentForLineno{#1*}%
}
\AtBeginDocument{%
\patchBothAmsMathEnvironmentsForLineno{equation}%
\patchBothAmsMathEnvironmentsForLineno{align}%
\patchBothAmsMathEnvironmentsForLineno{flalign}%
\patchBothAmsMathEnvironmentsForLineno{alignat}%
\patchBothAmsMathEnvironmentsForLineno{gather}%
\patchBothAmsMathEnvironmentsForLineno{multline}%
}

\usepackage{hyperref}    
\usepackage[all]{hypcap} 

\def\lhcb {\mbox{LHCb}\xspace}
\def\ux85 {\mbox{UX85}\xspace}

\ifthenelse{\boolean{uprightparticles}}%
{

 \def\Ppsi        {\ensuremath{\uppsi}\xspace}

 \def\PDelta      {\ensuremath{\Delta}\xspace}                 
 \def\PXi      {\ensuremath{\Xi}\xspace}                 
 \def\PLambda      {\ensuremath{\Lambda}\xspace}                 
 \def\PSigma      {\ensuremath{\Sigma}\xspace}                 
 \def\POmega      {\ensuremath{\Omega}\xspace}                 
 \def\PUpsilon      {\ensuremath{\Upsilon}\xspace}                 
 
 \def\PB      {\ensuremath{\mathrm{B}}\xspace}                 
                  
 \def\PD      {\ensuremath{\mathrm{D}}\xspace}

 \def\PJ      {\ensuremath{\mathrm{J}}\xspace}                 
 \def\PK      {\ensuremath{\mathrm{K}}\xspace}

 \def\Pb      {\ensuremath{\mathrm{b}}\xspace}

 \def\Pi      {\ensuremath{\mathrm{i}}\xspace}

 \def\Ps      {\ensuremath{\mathrm{s}}\xspace}

}
{

 \def\Ppsi        {\ensuremath{\psi}\xspace}                 
                  
 \mathchardef\PDelta="7101
 \mathchardef\PXi="7104
 \mathchardef\PLambda="7103
 \mathchardef\PSigma="7106
 \mathchardef\POmega="710A
 \mathchardef\PUpsilon="7107
                  
 \def\PB      {\ensuremath{B}\xspace}                 
                  
 \def\PD      {\ensuremath{D}\xspace}

 \def\PJ      {\ensuremath{J}\xspace}                 
 \def\PK      {\ensuremath{K}\xspace}

 \def\Pb      {\ensuremath{b}\xspace}

 \def\Pi      {\ensuremath{i}\xspace}

 \def\Ps      {\ensuremath{s}\xspace}

}

\def\squark    {\ensuremath{\Ps}\xspace}

\def\bquark    {\ensuremath{\Pb}\xspace}

\def\kaon  {\ensuremath{\PK}\xspace}
  \def\Kbar  {\kern 0.2em\overline{\kern -0.2em \PK}{}\xspace}

\def\Kz    {\ensuremath{\kaon^0}\xspace}
\def\Kzb   {\ensuremath{\Kbar^0}\xspace}
\def\KzKzb {\ensuremath{\Kz \kern -0.16em \Kzb}\xspace}
\def\Kp    {\ensuremath{\kaon^+}\xspace}
\def\Km    {\ensuremath{\kaon^-}\xspace}

\def\KpKm  {\ensuremath{\Kp \kern -0.16em \Km}\xspace}

\def\Kstarz  {\ensuremath{\kaon^{*0}}\xspace}
\def\Kstarzb {\ensuremath{\Kbar^{*0}}\xspace}
\def\Kstar   {\ensuremath{\kaon^*}\xspace}

  \def\Dbar    {\kern 0.2em\overline{\kern -0.2em \PD}{}\xspace}
\def\D       {\ensuremath{\PD}\xspace}

\def\Dz      {\ensuremath{\D^0}\xspace}
\def\Dzb     {\ensuremath{\Dbar^0}\xspace}
\def\DzDzb   {\ensuremath{\Dz {\kern -0.16em \Dzb}}\xspace}
\def\Dp      {\ensuremath{\D^+}\xspace}
\def\Dm      {\ensuremath{\D^-}\xspace}

\def\DpDm    {\ensuremath{\Dp {\kern -0.16em \Dm}}\xspace}

\def\B       {\ensuremath{\PB}\xspace}
  \def\Bbar    {\kern 0.18em\overline{\kern -0.18em \PB}{}\xspace}

\def\Bd      {\ensuremath{\B^0}\xspace}
\def\Bs      {\ensuremath{\B^0_\squark}\xspace}
\def\Bsb     {\ensuremath{\Bbar^0_\squark}\xspace}

\def\jpsi     {\ensuremath{{\PJ\mskip -3mu/\mskip -2mu\Ppsi\mskip 2mu}}\xspace}

  \def\Y#1S{\ensuremath{\PUpsilon{(#1S)}}\xspace}

\def\Lbar {\ensuremath{\kern 0.1em\overline{\kern -0.1em\PLambda}}\xspace}

\def\BF         {{\ensuremath{\cal B}\xspace}}

\def\to                 {\ensuremath{\rightarrow}\xspace}

\def\CP                {\ensuremath{C\!P}\xspace}

\def\AT#1     {\ensuremath{A_{\mathrm{T}}^{#1}}\xspace}           

\def\C#1      {\ensuremath{\mathcal{C}_{#1}}\xspace}                       
\def\Cp#1     {\ensuremath{\mathcal{C}_{#1}^{'}}\xspace}                    
\def\Ceff#1   {\ensuremath{\mathcal{C}_{#1}^{\mathrm{(eff)}}}\xspace}        
\def\Cpeff#1  {\ensuremath{\mathcal{C}_{#1}^{'\mathrm{(eff)}}}\xspace}       
\def\Ope#1    {\ensuremath{\mathcal{O}_{#1}}\xspace}                       
\def\Opep#1   {\ensuremath{\mathcal{O}_{#1}^{'}}\xspace}                    

\newcommand{\tev}{\ensuremath{\mathrm{\,Te\kern -0.1em V}}\xspace}
\newcommand{\gev}{\ensuremath{\mathrm{\,Ge\kern -0.1em V}}\xspace}
\newcommand{\mev}{\ensuremath{\mathrm{\,Me\kern -0.1em V}}\xspace}
\newcommand{\kev}{\ensuremath{\mathrm{\,ke\kern -0.1em V}}\xspace}
\newcommand{\ev}{\ensuremath{\mathrm{\,e\kern -0.1em V}}\xspace}
\newcommand{\gevc}{\ensuremath{{\mathrm{\,Ge\kern -0.1em V\!/}c}}\xspace}
\newcommand{\mevc}{\ensuremath{{\mathrm{\,Me\kern -0.1em V\!/}c}}\xspace}
\newcommand{\gevcc}{\ensuremath{{\mathrm{\,Ge\kern -0.1em V\!/}c^2}}\xspace}
\newcommand{\gevgevcccc}{\ensuremath{{\mathrm{\,Ge\kern -0.1em V^2\!/}c^4}}\xspace}
\newcommand{\mevcc}{\ensuremath{{\mathrm{\,Me\kern -0.1em V\!/}c^2}}\xspace}

\def\invfb   {\ensuremath{\mbox{\,fb}^{-1}}\xspace}

\def\gsim{{~\raise.15em\hbox{$>$}\kern-.85em
          \lower.35em\hbox{$\sim$}~}\xspace}
\def\lsim{{~\raise.15em\hbox{$<$}\kern-.85em
          \lower.35em\hbox{$\sim$}~}\xspace}

\def\pt         {\mbox{$p_{\rm T}$}\xspace}

\def\evtgen     {\mbox{\textsc{EvtGen}}\xspace}
\def\pythia     {\mbox{\textsc{Pythia}}\xspace}

\def\geant      {\mbox{\textsc{Geant4}}\xspace}

\def\gauss      {\mbox{\textsc{Gauss}}\xspace}

\def\photos     {\mbox{\textsc{Photos}}\xspace}

\def\tell1  {TELL1\xspace}
\def\ukl1   {UKL1\xspace}

\usepackage{cite} 
\usepackage{mciteplus}

\newcommand{\swave}{{S-wave}\xspace}
\newcommand{\pwave}{{P-wave}\xspace}
\newcommand{\dwave}{{D-wave}\xspace}
\newcommand{\fwave}{{F-wave}\xspace}
\newcommand{\gwave}{{G-wave}\xspace}

\newcommand{\Kst}{\ensuremath{K^{*0}}\xspace}
\newcommand{\BdJpsiKst}{\ensuremath{\Bd \to J/\psi K^{*0}}\xspace}
\newcommand{\BdJpsiKpi}{\ensuremath{\Bd \to J/\psi K \pi}\xspace}

\def\Kstarzma  {\ensuremath{\kaon^{*}(892)^0}\xspace}
\def\Kstarzbma {\ensuremath{\Kbar^{*}(892)^0}\xspace}

\newcommand{\BsJpsiKstma}{\ensuremath{B^0_s\to J/\psi \Kstarzbma}\xspace}

\newcommand{\BJpsiKpi}{\ensuremath{B^0_{(s)}\to J/\psi K \pi}\xspace}
\newcommand{\BsJpsiKst}{\ensuremath{B^0_s\to J/\psi \Kstarzb}\xspace}
\newcommand{\BsJpsiKpi}{\ensuremath{B^0_s\to J/\psi K \pi}\xspace}
\newcommand{\fdfs}{\ensuremath{\frac{f_d}{f_s}}\xspace}
\newcommand{\BdJpsiRho}{\ensuremath{B^0_d\to J/\psi \rho^0}\xspace}

\newcommand{\BRof}[1]{\ensuremath{{\cal B}(#1)}\xspace}

\newcommand{\BuJpsiK}{\ensuremath{B^+\to J/\psi K^+}\xspace}

\newcommand{\IP}{\ensuremath{{\rm IP}}\xspace}
\newcommand{\pdf}{\ensuremath{\rm PDF}\xspace}

\newcommand{\figref}[1]{Fig.~\ref{#1}}
\newcommand{\tabref}[1]{Table~\ref{#1}}

\usepackage{accents}
\newcommand*{\fancybar}{\scalebox{.4}{(}\raisebox{-1.7pt}{\bf{--}}\scalebox{.4}{)}}
\newcommand*{\brabar}[1]{\accentset{\fancybar}{#1}}
\newcommand{\BJpsiKst}{\ensuremath{\Bd_{(s)}\to J/\psi \brabar{K}^{*0}}\xspace}

\begin{document}

\renewcommand{\thefootnote}{\fnsymbol{footnote}}
\setcounter{footnote}{1}





\begin{titlepage}
\pagenumbering{roman}

\vspace*{-1.5cm}
\centerline{\large EUROPEAN ORGANIZATION FOR NUCLEAR RESEARCH (CERN)}
\vspace*{1.5cm}
\hspace*{-0.5cm}
\begin{tabular*}{\linewidth}{lc@{\extracolsep{\fill}}r}
\ifthenelse{\boolean{pdflatex}}
{\vspace*{-2.7cm}\mbox{\!\!\!\includegraphics[width=.14\textwidth]{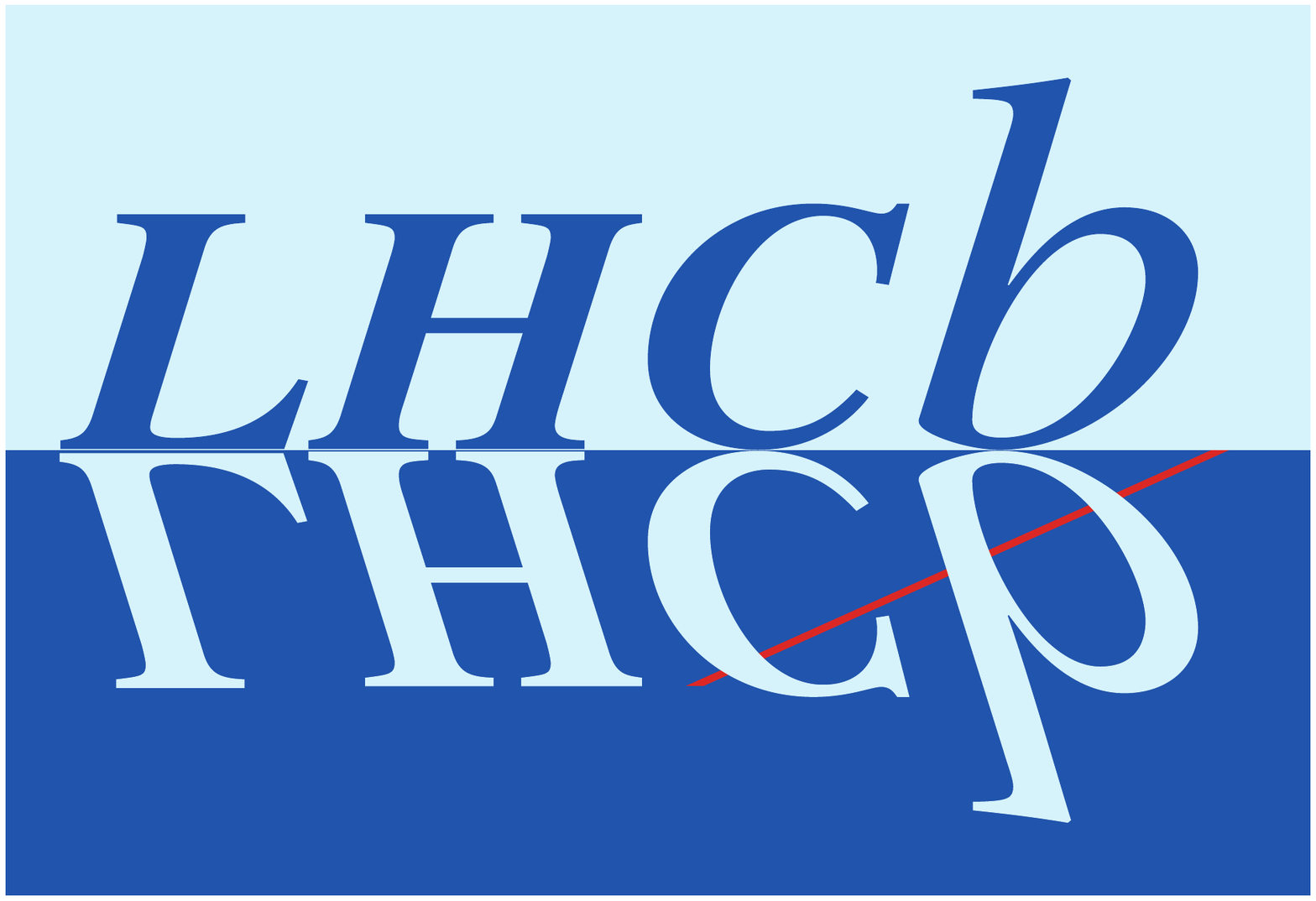}} & &}%
{\vspace*{-1.2cm}\mbox{\!\!\!\includegraphics[width=.12\textwidth]{lhcb-logo.eps}} & &}%
\\
 & & CERN-PH-EP-2012-214 \\  
 & & LHCb-PAPER-2012-014 \\  
 & & September 20, 2012 \\ 
 & & \\
\end{tabular*}

\vspace*{3.0cm}

{\bf\boldmath\huge
\begin{center}
  Measurement of the \BsJpsiKst branching fraction and angular amplitudes
\end{center}
}

\vspace*{2.0cm}

\begin{center}
The LHCb collaboration\footnote{Authors are listed on the following pages.}
\end{center}

\vspace{\fill}


\begin{abstract}

\noindent A sample of $114 \pm 11$ $\Bs \to \jpsi K^-\pi^+$ signal events
obtained with \mbox{$0.37$ fb$^{-1}$} of $pp$ collisions at $\sqrt{s}$ = 7~TeV
collected by the LHCb experiment is used to measure the branching fraction and
polarization amplitudes of the \BsJpsiKst decay, with $\Kstarzb \to K^-\pi^+$.
The $K^-\pi^+$ mass spectrum of the candidates in the \Bs peak is dominated by
the \Kstarzb contribution. Subtracting the non-resonant $K^-\pi^+$ component,
the branching fraction of \BsJpsiKst is $\left( 4.4_{-0.4}^{+0.5} \pm 0.8
\right) \times 10^{-5}$, where the first uncertainty is statistical and the
second is systematic. A fit to the angular distribution of the decay products
yields the \Kst polarization fractions $f_L = 0.50 \pm 0.08 \pm 0.02$ and
$f_{\parallel} = 0.19^{+0.10}_{-0.08} \pm 0.02$.

\end{abstract}

\vspace*{1.0cm}

\begin{center}
Submitted to Physical Review D (R)
\end{center}

\vspace{\fill}

\end{titlepage}


\newpage
\setcounter{page}{2}
\mbox{~}
\newpage


\centerline{\large\bf LHCb collaboration}
\begin{flushleft}
\small
R.~Aaij$^{38}$, 
C.~Abellan~Beteta$^{33,n}$, 
A.~Adametz$^{11}$, 
B.~Adeva$^{34}$, 
M.~Adinolfi$^{43}$, 
C.~Adrover$^{6}$, 
A.~Affolder$^{49}$, 
Z.~Ajaltouni$^{5}$, 
J.~Albrecht$^{35}$, 
F.~Alessio$^{35}$
% ...

{\footnotesize \it
$ ^{1}$Centro Brasileiro de Pesquisas F\'{i}sicas (CBPF), Rio de Janeiro, Brazil\\
% ...
$ ^{5}$Clermont Universit\'{e}, Universit\'{e} Blaise Pascal, CNRS/IN2P3, LPC, Clermont-Ferrand, France\\
$ ^{6}$CPPM, Aix-Marseille Universit\'{e}, CNRS/IN2P3, Marseille, France\\
% ...
$ ^{11}$Physikalisches Institut, Ruprecht-Karls-Universit\"{a}t Heidelberg, Heidelberg, Germany\\
% ...
$ ^{33}$Universitat de Barcelona, Barcelona, Spain\\
$ ^{34}$Universidad de Santiago de Compostela, Santiago de Compostela, Spain\\
$ ^{35}$European Organization for Nuclear Research (CERN), Geneva, Switzerland\\
% ...
$ ^{38}$Nikhef National Institute for Subatomic Physics, Amsterdam, The Netherlands\\
% ...
$ ^{43}$H.H. Wills Physics Laboratory, University of Bristol, Bristol, United Kingdom\\
% ...
$ ^{49}$Oliver Lodge Laboratory, University of Liverpool, Liverpool, United Kingdom\\
% ...
$ ^{n}$LIFAELS, La Salle, Universitat Ramon Llull, Barcelona, Spain\\
% ...
}
\end{flushleft}


\cleardoublepage



\renewcommand{\thefootnote}{\arabic{footnote}}
\setcounter{footnote}{0}



\pagestyle{plain} 
\setcounter{page}{1}
\pagenumbering{arabic}


%



\begin{figure}[b]
\centering
\includegraphics[width =0.45\textwidth]{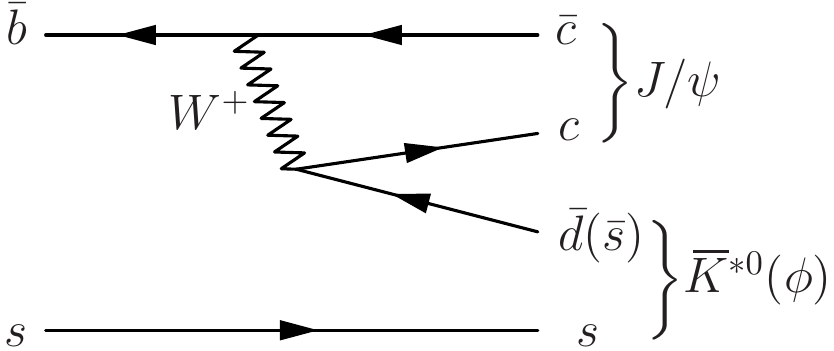}
\hspace{5mm}
\includegraphics[width =0.45\textwidth]{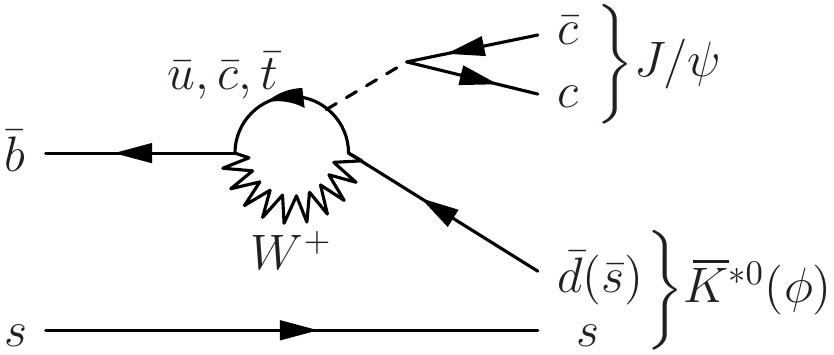}
\caption{\small Tree and penguin decay topologies contributing to the decays
\BsJpsiKst and $\Bs\to \jpsi\phi$. The dashed line indicates a colour singlet
exchange.}
\label{fig:feynman}
\end{figure}

Interpretations of measurements of time-dependent \CP violation in $\Bs\to
\jpsi\phi$ and $\Bs \to \jpsi f_0(980)$ decays have thus far assumed the
dominance of the colour-suppressed tree-level process. However, there are
contributions from higher order (penguin) processes (see
Fig.~\ref{fig:feynman}) that cannot be calculated reliably in QCD and could be
large enough to affect the measured asymmetries. It has been suggested that the
penguin effects can be determined by means of an analysis of the angular
distribution of $\BsJpsiKstma$, where the penguin diagram is not suppressed
relative to the tree-level one, and $SU(3)$ flavour symmetry arguments can be
used to determine the hadronic parameters entering the $\Bs\to \jpsi\phi$
observables~\cite{faller}.

In this paper the \Kstarzma meson will be written as \Kstarz, while for other
$K^*$ resonances the mass will be given in parentheses. Furthermore, mention of
any specific mode implies the use of the charge conjugated mode as well, and
$K^{-}\pi^{+}$ pairs will be simply written as $K\pi$. The decay \BsJpsiKst has
already been observed by the CDF experiment~\cite{CDF}, which reported
$\BRof\BsJpsiKst = (8.3\pm3.8)\times10^{-5}$.
Under the assumption that the light quark
($s$,$d$) is a spectator of the $b$ quark decay, the branching fraction can be
approximated as
\begin{equation}
\BRof\BsJpsiKst \sim \frac{|V_{cd}|^2}{|V_{cs}|^2} \times \BRof\BdJpsiKst
 =  (6.5 \pm 1.0) \times10^{-5},
\label{eq:theory}
\end{equation}
with $|V_{cd}| = 0.230 \pm 0.011$, $|V_{cs}| = 1.023 \pm 0.036$ \cite{PDG}, and
$\BRof\BdJpsiKst = (1.29 \pm 0.05 \pm 0.13) \times 10^{-3}$ \cite{Belle}. The
measurement in Ref.~\cite{Belle}, where the $K\pi$ \swave contribution is
subtracted, is used instead of the PDG average.

In this paper, $0.37 \invfb$ of data taken in 2011 are used to determine
\BRof\BsJpsiKst, to study the angular properties of the decay products of the
\Bs meson, and to measure the resonant contributions to the $K\pi$ spectrum in
the region of the \Kstarz meson. The measurement of the branching fraction uses
the decay \BdJpsiKst as a normalization mode.


The \lhcb detector~\cite{Alves:2008zz} is a single-arm forward spectrometer
covering the pseudo-rapidity range \mbox{$2<\eta<5$}. The detector includes a
high precision tracking system consisting of a silicon-strip vertex detector
located around the interaction point, a large-area silicon-strip detector
located upstream of a dipole magnet with a bending power of about $4{\rm\,Tm}$,
and three stations of silicon-strip detectors and straw drift tubes placed
downstream. The combined tracking system has a momentum resolution $\Delta p/p$
that varies from 0.4\,\% at 5\gevc to 0.6\,\% at 100\gevc. Two ring-imaging
Cherenkov detectors (RICH) are used to determine the identity of charged
particles. The separation of pions and kaons is such that, for efficiencies of $\sim75\%$ 
the rejection power is above $99\%$.
 Photon, electron and hadron candidates are identified by a
calorimeter system consisting of scintillating-pad and pre-shower detectors, an
electromagnetic calorimeter and a hadronic calorimeter. Muons are identified by
alternating layers of iron and multiwire proportional chambers.

The trigger consists of a hardware stage, based on information from the
calorimeter and muon systems, followed by a software stage called High Level
Trigger (HLT) that applies a full event reconstruction. Events with muon final
states are triggered using two hardware trigger decisions: the single-muon
decision (one muon candidate with transverse momentum $p_{\rm T}> 1.5$\gevc),
and the di-muon decision (two muon candidates with $p_{{\rm T},1}$ and $p_{{\rm
T},2}$ such that $\sqrt{p_{{\rm T},1} \, p_{{\rm T},2}} > 1.3$\gevc). All
tracks in the HLT are required to have a $p_{\rm T}> 0.5$\gevc.  The single
muon trigger decision in the HLT selects events with at least one muon track
with an impact parameter \IP$>0.1$\,mm with respect to the primary vertex and
$p_{\rm T}>1.0$\gevc. The di-muon trigger decision, designed to select $J/\psi$
mesons, also requires a di-muon mass ($M_{\mu\mu}$) $2970 < M_{\mu\mu} <
3210$\mevcc.

Simulated events are used to compute detection efficiencies and angular
acceptances. For this purpose, $pp$ collisions are generated using
\pythia~6.4~\cite{Sjostrand:2006za} with a specific \lhcb
configuration~\cite{LHCb-PROC-2010-056}.  Decays of hadronic particles are
described by \evtgen~\cite{Lange:2001uf} in which final state radiation is
generated using \photos~\cite{Golonka:2005pn}. The interaction of the generated
particles with the detector and its response are implemented using the \geant
toolkit~\cite{Allison:2006ve, *Agostinelli:2002hh} as described in
Ref.~\cite{LHCb-PROC-2011-006}.

The selection of \BJpsiKst decays first requires the reconstruction of a $\jpsi
\to \mu^+\mu^-$ candidate. The \jpsi vertex is required to be separated from 
any primary vertex (PV) by a distance-of-flight significance greater than 13.
Subsequently, the muons from the \jpsi decay are combined with the $K$ and
$\pi$ candidates to form a good vertex, where the di-muon mass is constrained
to the \jpsi mass. A $\pt > 0.5$\gevc is required for each of the four
daughter tracks. Positive muon identification is required for the two tracks of
the \jpsi decay, and the kaons and pions are selected using the different
hadron probabilities based on combined information given by the RICH detectors.
The candidate $B^0_{(s)}$ momentum is required to be compatible with the flight
direction as given by the vector connecting the PV with the candidate vertex.
An explicit veto to remove \BuJpsiK events is applied, as they otherwise would
pollute the upper sideband of the $B^0_{(s)}$ mass spectrum.

Following this initial selection, several geometrical variables are combined
into a single discriminant geometrical likelihood variable (GL). This
multivariate method is described in Refs.~\cite{DiegoThesis, Karlen}. The
geometrical variables chosen to build the GL are: the $B^0_{(s)}$ candidate
minimum impact parameter with respect to any PV in the event, the decay time of
the $B^0_{(s)}$ candidate, the minimum impact parameter $\chi^2$ of the four
daughter tracks with respect to all PV in the event (defined as the difference
between the $\chi^2$ of the PV built with and without the considered track),
the distance of closest approach between the \jpsi and \Kstarz trajectories
reconstructed from their decay products, and the $\pt$ of the $B^0_{(s)}$
candidate. The GL was tuned using simulated \BdJpsiKst signal passing the
selection criteria, and background from data in the $B^0_{(s)}$ mass sidebands
with a value for the kaon particle identification variable in a range which
does not overlap with the one used to select the data sample for the final
analysis. 

\begin{figure}[t]
\centering
\includegraphics[width =0.48\textwidth]{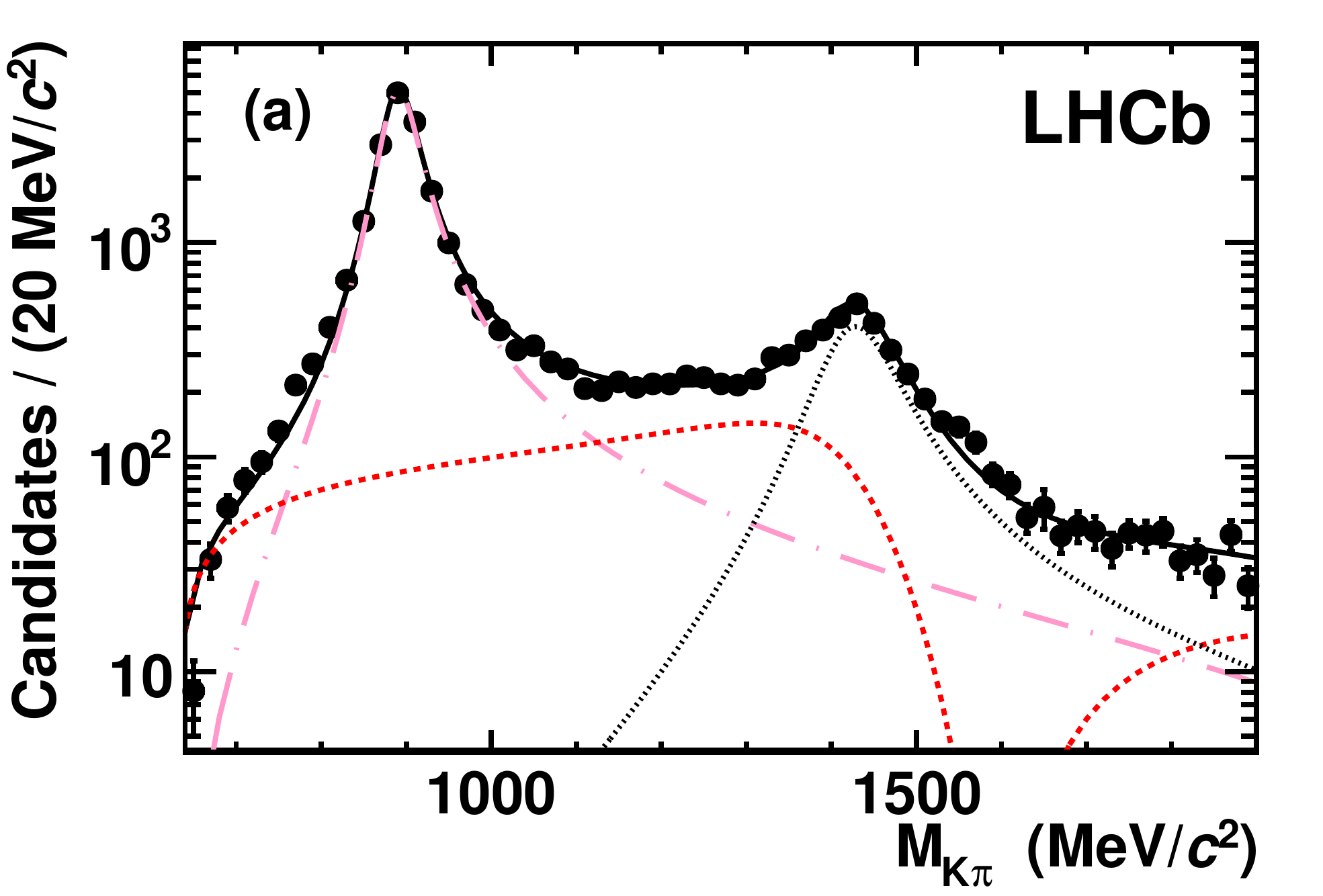}
\includegraphics[width =0.48\textwidth]{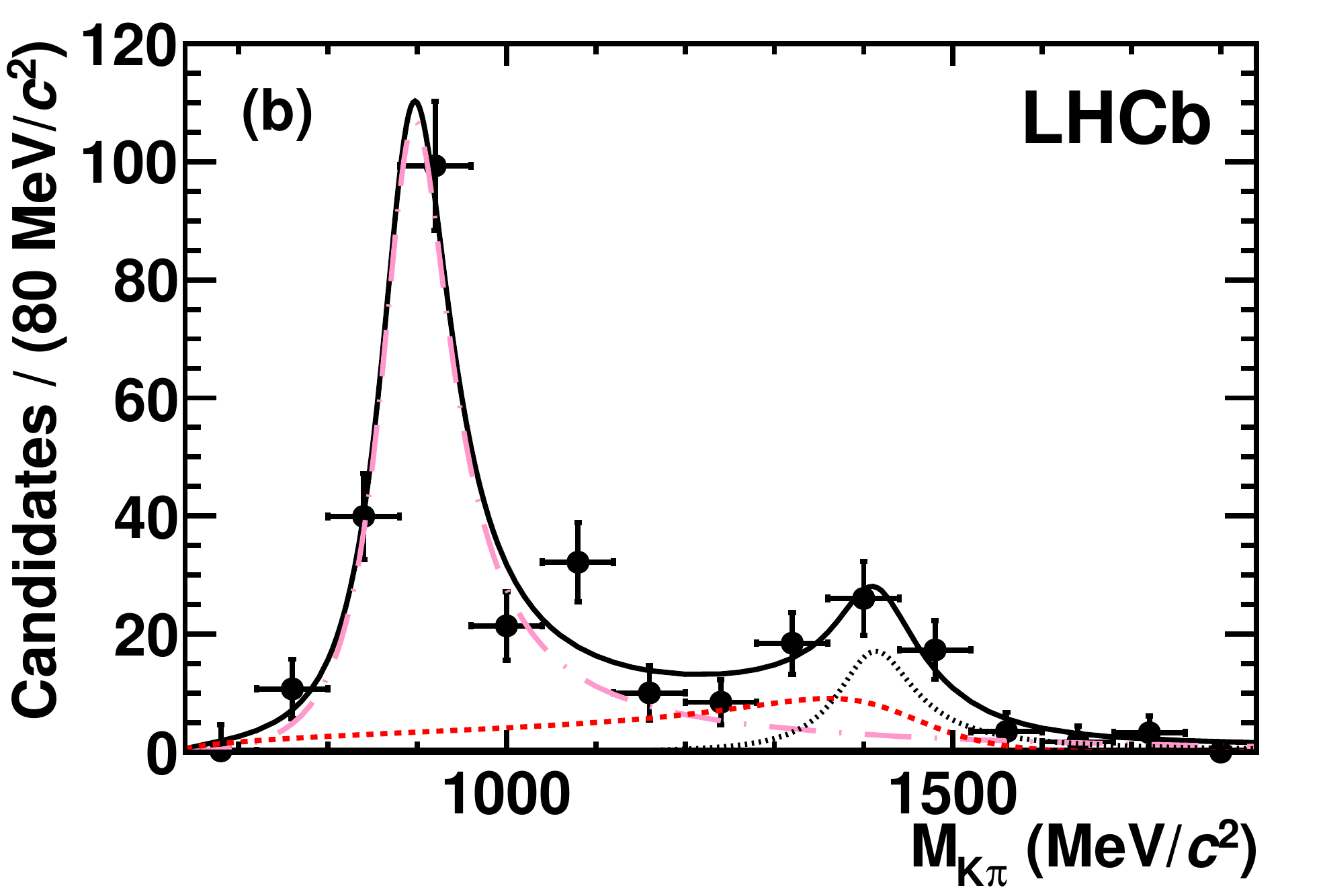}
\caption{\small Fit to the $K\pi$ mass spectrum for (a) \BdJpsiKpi events, and
(b) \BsJpsiKpi events. The \BJpsiKpi yields in each bin
of $K\pi$ mass are determined from a fit to the
$\jpsi K \pi$ mass spectrum. The pink dashed-dotted line represents the
\Kstarz, the red short-dashed line is the \swave and the black dotted line is
the $K^{*}_2(1430)$. The black solid line is their sum.}
\label{fig:Kpi_TwoBW}
\end{figure}

The $K\pi$ mass spectrum in the \BdJpsiKpi channel is dominated by the \Kstarz
resonance but contains a non-negligible \swave contribution, originating from
$K^{*}_0(1430)^0$ and non-resonant $K\pi$ pairs~\cite{z4430}. To determine
\BRof\BsJpsiKst it is therefore important to measure the \swave magnitude in
both $\Bd_{(s)} \to \jpsi K\pi$ channels. The $K\pi$ spectrum is analyzed in
terms of a non-resonant \swave and several $K\pi$ resonances parameterized
using relativistic Breit-Wigner distributions with mass-dependent widths,
following closely ~\cite{z4430}.  The considered waves are: a non-resonant
\swave amplitude interfering with the $K^{*}_0(1430)^0$ resonance, \Kstarz for
the \pwave and $K^{*}_2(1430)^0$ for the \dwave.  \fwave and \gwave components
are found to be negligible in the \Bd fit. In bins of the $K\pi$ mass, a fit is
made to the $B^0_{(s)}$ candidate mass distribution to determine the yield. As
shown in \figref{fig:Kpi_TwoBW}, a fit is then made to the \Bd and \Bs yields
as a function of the  $K\pi$ mass without any efficiency correction. The S and
\pwave components dominate in the $\pm 40\, \mevcc$ window around the \Kstarz
mass, where the \Kstarz contribution is above 90\%. A more exact determination
of this contribution using this method would require $K\pi$ mass-dependent
angular acceptance corrections. For the branching fraction calculation, the
fraction of \Kstarz candidates is determined from a different full angular and
mass fit, which is described next.

The angular and mass analysis is based on an unbinned maximum likelihood fit
which handles simultaneously the mass ($M_{\jpsi K \pi}$) and the angular
parameters of the $B^0_{(s)}$ decays and the background. Each of these three
components is is modelled as a product of probability density functions (\pdf),
${\cal P}(M_{\jpsi K\pi}, \psi, \theta,\varphi) = {\cal P}(M_{\jpsi K\pi})\,
{\cal P}( \psi, \theta,\varphi),$ with $\psi$ the angle between the kaon
momentum in the rest frame of the \Kstarz and the direction of motion of the
\Kstarz in the rest frame of the \B. The polar and azimuthal angles ($\theta$,
$\varphi$) describe the direction of the $\mu^+$ in the coordinate system
defined in the \jpsi rest frame, where the $x$ axis is the direction of motion
of the $B^0_{(s)}$ meson, the $z$ axis is normal to the plane formed by the $x$
axis and the kaon momentum, and the $y$ axis is chosen so that the $y$
component of the kaon momentum is positive. 

The function describing the mass distribution of both $B^0_{(s)}$ signal peaks
is the sum of two Crystal Ball (CB) functions~\cite{Skwarnicki:1986xj}, which
are a combination of a Gaussian and a power law function to describe the
radiative tail at low masses,
\begin{equation}
{\cal P}(M_{\jpsi K\pi}) = f \, {\rm CB}(M_{\jpsi K\pi},\mu_{\B}, \sigma_1,
 \alpha_1) + 
(1-f)\, \operatorname{CB}(M_{\jpsi K\pi}, \mu_{\B}, \sigma_2, \alpha_2).
\end{equation}
\noindent The starting point of the radiative tail is governed by a transition
point parameter $\alpha_{(1,2)}$. The mean and width of the Gaussian component
are $\mu_{\B}$ and $\sigma_{(1,2)}$.  The values of the $f$, $\sigma_1$,
$\sigma_2$, $\alpha_1$ and $\alpha_2$ parameters are constrained to be the same
for the \Bs and \Bd peaks. The difference in the means between the \Bs and the
$\Bd$ distributions, ($\mu_{\Bs} - \mu_{\Bd}$), is fixed to the value taken
from Ref.~\cite{LHCb-PAPER-2011-035}. The mass \pdf of the background is
described by an exponential function.

Assuming that direct \CP violation and the $\B^0_{(s)} - \Bbar^0_{(s)}$
production asymmetry are insignificant, the differential decay rate
is~\cite{faller,babar_ang}

\begin{eqnarray}
\label{eq:angpdf}
\frac{{\rm d}^3\Gamma}{{\rm d}\Omega} &\propto&
2 |A_0|^2 \cos^{2}\psi(1-\sin^{2}\theta \cos^{2}\varphi) \nonumber \\
&+& |A_{\parallel}|^2 \sin^{2}\psi(1-\sin^{2}\theta \sin^{2}\varphi)
 \nonumber \\
&+& |A_{\perp}|^2 \sin^{2}\psi \sin^{2}\theta  \nonumber \\
&+& \frac{1}{\sqrt{2}} |A_{0}| |A_{\parallel}| \cos(\delta_{\parallel}-
\delta_0) \sin 2\psi \sin^2\theta \sin 2\varphi \\
&+& \frac{2}{3}|A_{\rm S}|^2  \left[ 1 - \sin^2\theta \cos^2\varphi \right]
\nonumber \\
&+& \frac{4\sqrt{3}}{3}|A_{0}| |A_{\rm S}| \cos(\delta_{\rm S}-\delta_0)
  \cos\psi \left[1 -\sin^2\theta \cos^2\varphi  \right] \nonumber \\
&+& \frac{\sqrt{6}}{3} |A_{\parallel}| |A_{\rm S}| \cos(\delta_{\parallel}-
\delta_{\rm S}) \sin\psi \sin^2\theta \sin2\varphi,\nonumber
\end{eqnarray}
where $A_0$, $A_{\parallel}$ and $A_{\perp}$ are the decay amplitudes
corresponding to longitunally and transversely polarized vector mesons. $A_{\rm
S}=|A_{\rm S}|e^{i\delta_{\rm S}}$ is the $K\pi$ \swave amplitude and
($\delta_{\parallel} - \delta_0$) the relative phase between the longitudinal
and parallel amplitudes. The convention $\delta_0 = 0$ is used hereafter. The
$\Omega$ differential is ${\rm d}\Omega \equiv {\rm d}\cos\psi\, {\rm
d}\cos\theta\, {\rm d}\varphi$. The polarization fractions are normalized
according to
\begin{equation} f_{L,\parallel,\perp} =
\frac{|A_{0,\parallel,\perp}|^2}{|A_0|^2 + |A_{\parallel}|^2 +|A_{\perp}|^2},
\end{equation} which satisfy $f_L + f_{\parallel}+f_{\perp} = 1$.

The parameters $f_L$, $f_{\parallel}$ and $\delta_{\parallel}$ describing the
\pwave are left floating in the fit. The $|A_{\rm S}|$ amplitude and the
$\delta_{\rm S}$ phase depend on $M_{K \pi}$, but this dependence is ignored in
the fit, which is performed in a $K\pi$ mass window of $\pm 40 \mevcc$, and
they are just treated also as floating parameters. A systematic uncertainty is
later associated with this assumption. The angular distribution of observed
events is parameterized as a product of the expression in Eq.~\ref{eq:angpdf}
and a detector acceptance function, ${\rm Acc}(\Omega)$, which describes the
efficiency to trigger, reconstruct and select the events. Simulation studies
have shown almost no correlation between the three one-dimensional angular
acceptances ${\rm Acc}_{\psi}(\psi)$, ${\rm Acc}_{\theta}(\theta)$ and ${\rm
Acc}_{\varphi}(\varphi)$. Therefore the global acceptance factorizes as ${\rm
Acc}(\Omega) = {\rm Acc}_{\psi}(\psi)\, {\rm Acc}_{\theta}(\theta)\, {\rm
Acc}_{\varphi}(\varphi)$, where ${\rm Acc}_{\psi}(\psi)$ is parameterized as a
fifth degree polynomial, ${\rm Acc}_{\theta}(\theta)$ as a second degree
polynomial and ${\rm Acc}_{\phi}(\phi)$ as a sinusoidal function. A systematic
uncertainty due to this factorization hypothesis is later evaluated. The
angular distribution for the background component is determined using the upper
sideband of the \Bs mass spectrum, defined as the interval $[5417, 5779]$
\mevcc.

\begin{figure*}[t]
\begin{center}
\includegraphics[width=0.48\textwidth]{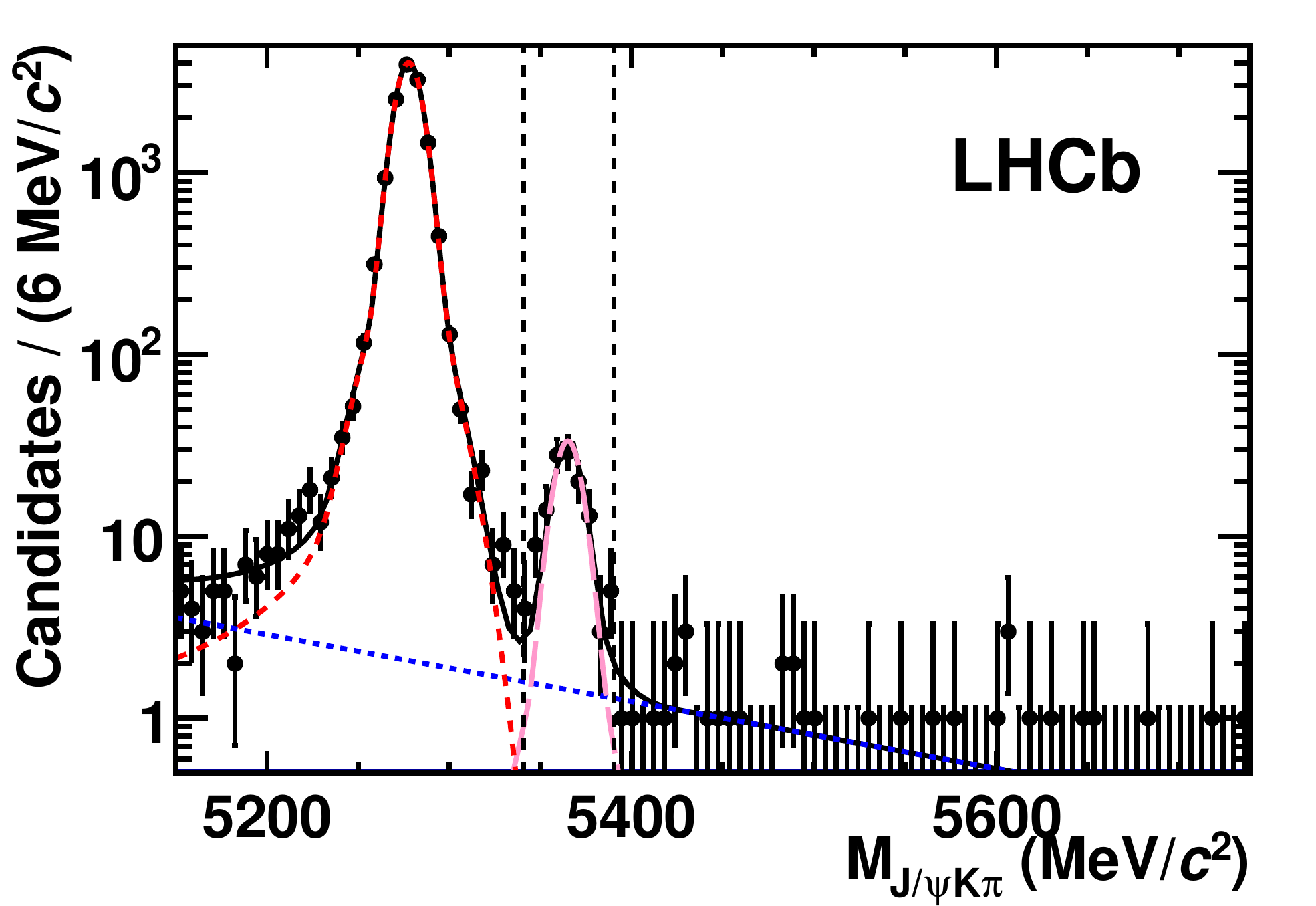}
\includegraphics[width=0.48\textwidth]{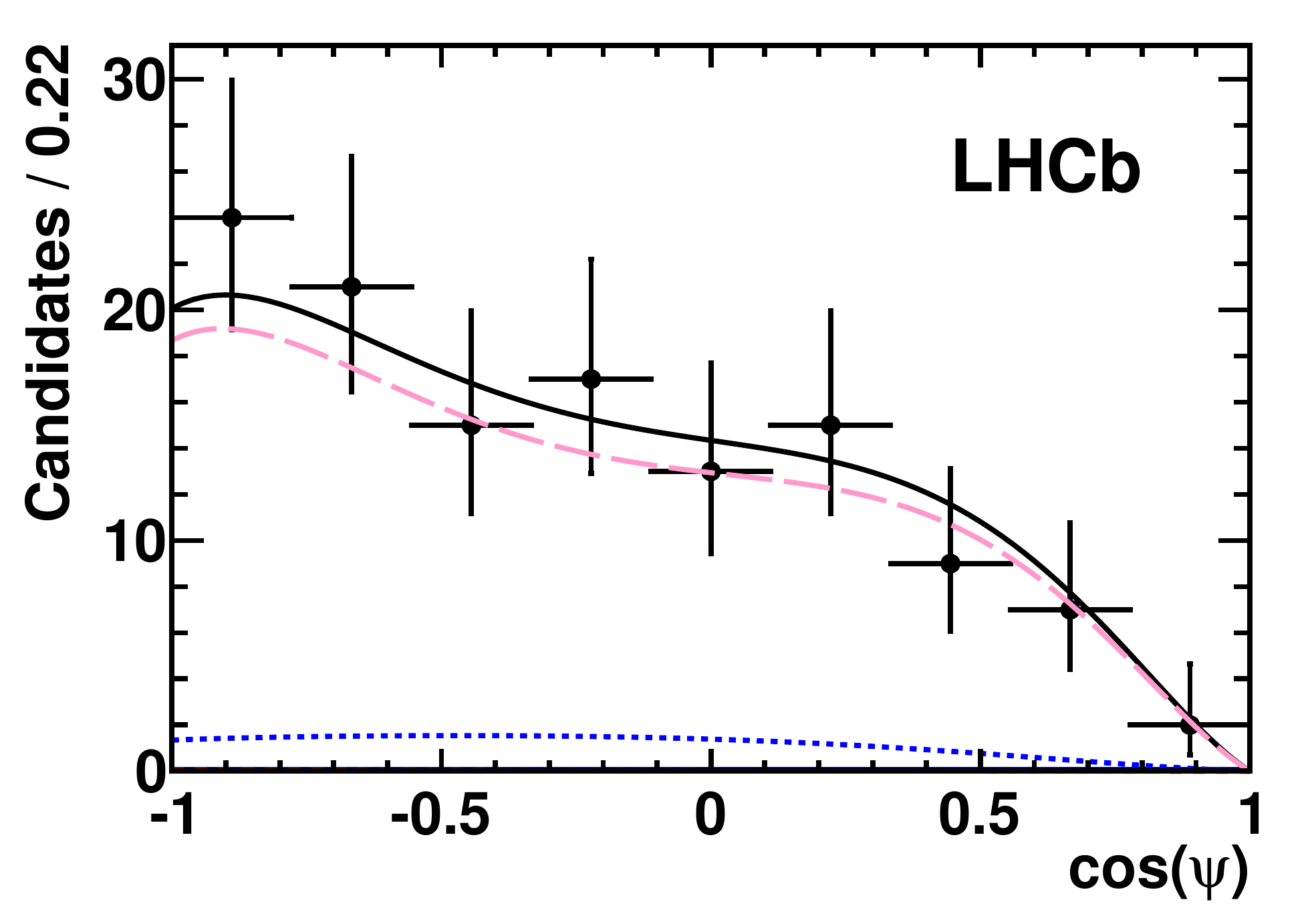}
\includegraphics[width=0.48\textwidth]{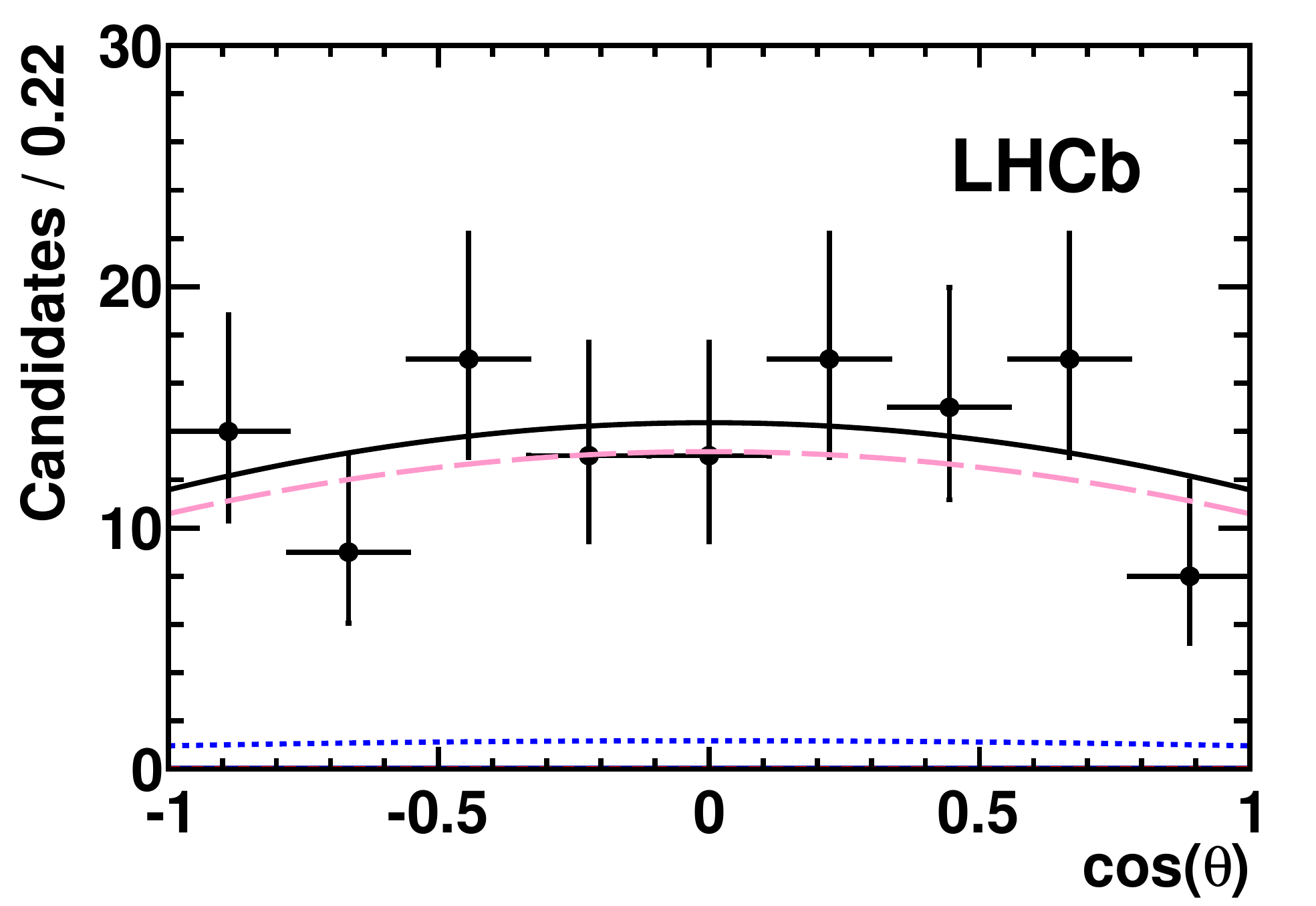}
\includegraphics[width=0.48\textwidth]{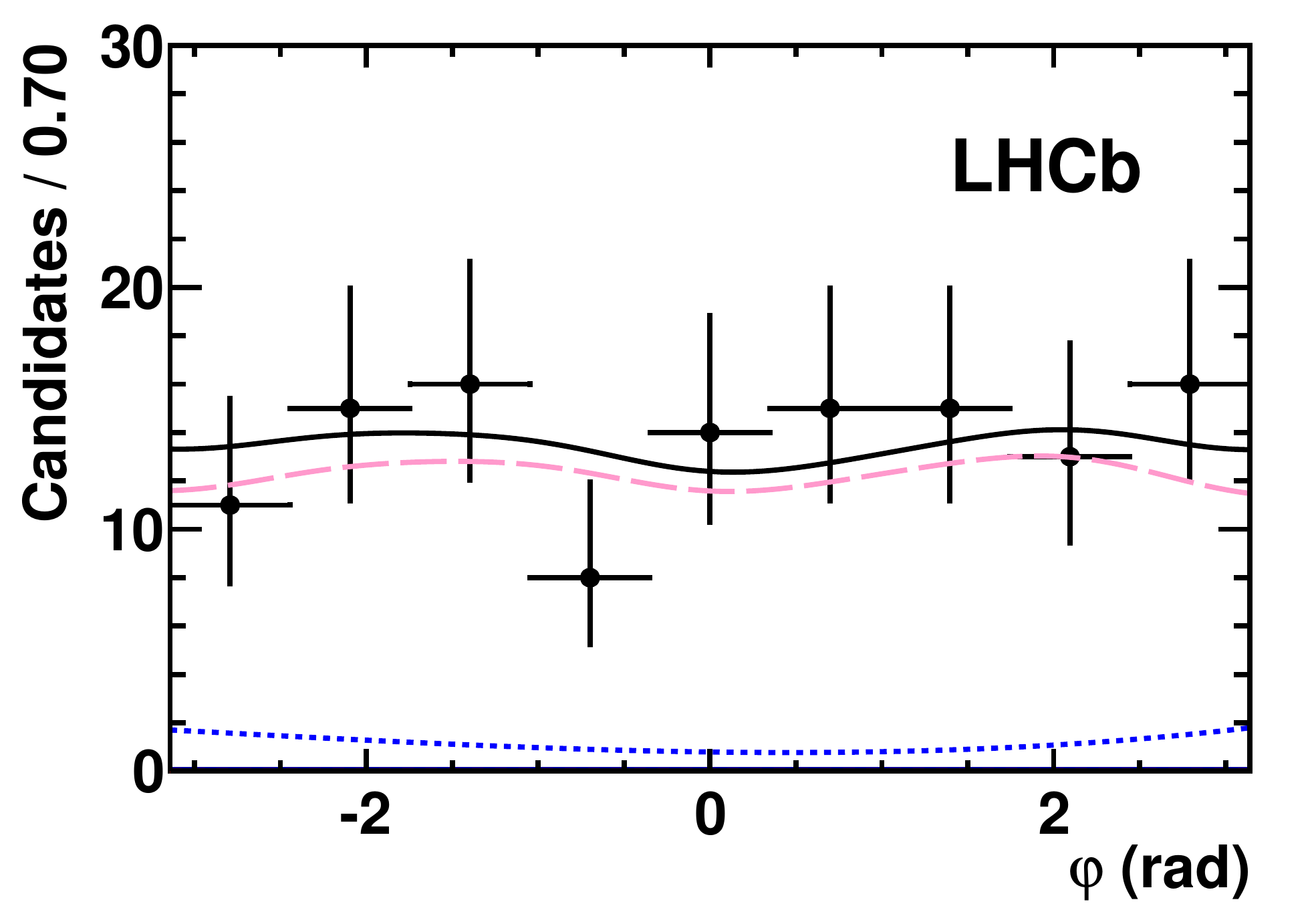}
\end{center}
\caption{\small Projections of the fit in $M_{J/\psi K\pi}$ and in the angular
variables for the mass range indicated by the two dashed vertical lines in the
mass plot. The red dashed, pink long-dashed, and blue dotted lines represent
the fitted contributions from \BdJpsiKst, \BsJpsiKst and background. The black
solid line is their sum.}
\label{fig:full_fit}
\end{figure*}

Figure \ref{fig:full_fit} shows the projection of the fit in the $M_{\jpsi
K\pi}$ mass axis, together with the projections in the angular variables in a
window of $\pm 25$ \mevcc around the \Bs mass. The number of candidates
corresponding to $\Bd$ and \Bs decays is found to be $13,\!365 \pm 116$ and
$114 \pm 11$, respectively.

\begin{center}
\begin{table*}[htb]
\centering
\caption[]{\small Summary of the measured \BsJpsiKst angular properties and
their statistical and systematic uncertainties.}
\vspace{3mm}
\centering
\setlength{\extrarowheight}{3pt}
\begin{tabular}{c|c|c|c}
Parameter name & $|A_{\rm S}|^2$ & $f_L$  & $f_{\parallel}$  \\ [+0.04in]
\hline
Value and statistical error & $0.07_{-0.07}^{+0.15} $  &  $0.50 \pm 0.08$ &
$0.19_{-0.08}^{+0.10}$ \\ [+0.03in] \hline
\multicolumn{4}{c}{Systematic uncertainties}\\
\hline
Angular acceptance      & $0.044$ & $0.011$  & $0.016$  \\
Background angular model                  & $0.038$ & $0.017$  & $0.013$  \\
Assumption $\delta_{\rm S}(M_{K\pi}) =$ constant & $0.026$ & $0.005$ & $0.002$\\
\Bd contamination      & $0.036$ & $0.004$ & $0.007$ \\
Fit bias                                 & $-$     & $-$      & $0.005$ \\
\hline
Total systematic error                    & $0.073$ &  $0.021$ & $0.022$ \\
\end{tabular}
\label{tab:final_Ang}
\end{table*}
\end{center}

\begin{center}
\begin{table*}[htb]
\centering
\caption[]{\small Angular parameters of \BdJpsiKst needed to compute
\BRof\BsJpsiKst. The systematic uncertainties from background modelling and the
mass \pdf are found to be negligible in this case.}
\vspace{3mm}
\centering
\setlength{\extrarowheight}{2pt}
\begin{tabular}{c|c|c|c}
Parameter name & $|A_{\rm S}|^2$ & $f_L$  & $f_{\parallel}$\\
\hline
Value and statistical error & $0.037 \pm 0.010$  &  $0.569 \pm 0.007$& $0.240
 \pm 0.009$ \\[+0.03in] \hline
\multicolumn{4}{c}{Systematic uncertainties}\\
\hline
Angular acceptance    & $0.044$ & $0.011$  & $0.016$  \\
Assumption $\delta_{\rm S}(M_{K\pi}) =$ constant & $0.026$ & $0.005$ & $0.002$\\
\hline
Total systematic error                    & $0.051$ &  $0.012$ & $0.016$ \\
\end{tabular}
\label{tab:final_Ang_Bd}
\end{table*}
\end{center}

Tables \ref{tab:final_Ang} and \ref{tab:final_Ang_Bd} summarize the
measurements of the \BJpsiKst angular parameters, together with their
statistical and systematic uncertainties. The correlation coefficient given by
the fit between $f_L$ and $f_\parallel$ is $\rho = -0.44$ for \Bs decays. The
results for the \BdJpsiKst decay are in good agreement with previous
measurements \cite{babar_ang,Belle,jpsikstar_pol3,jpsikstar_pol4}. Based on
this agreement, the systematic uncertainties caused by the modelling of the
angular acceptance were evaluated by summing in quadrature the statistical
error on the measured \BdJpsiKst parameters with the uncertainties on the world
averages ($f_L = 0.570 \pm 0.008$ and $f_{\perp}=0.219 \pm 0.010$)~\cite{PDG}.
The angular analysis was repeated with two additional acceptance descriptions,
one which uses a three-dimensional histogram to describe the efficiency
avoiding any factorization hypothesis, and another one based on a method of
normalization weights described in Ref.~\cite{tristan}. A very good agreement
was found in the values of the polarization fractions computed with all the
three methods.  For the parameter $|A_{\rm S}|^2$, uncertainties caused by the
finite size of the simulation sample used for the acceptance description, as
well as from the studies with several acceptance models, are included. The
systematic uncertainty caused by the choice of the angular \pdf for the
background is shown for the \BsJpsiKst decay but it was found to be negligible
for \BdJpsiKst. 

Also included in Tables \ref{tab:final_Ang} and \ref{tab:final_Ang_Bd} is the
uncertainty from the assumption of a constant $\delta_{\rm S}$ as a function of
$M_{K\pi}$. This assumption can be relaxed by adding an extra free parameter to
the angular \pdf. This addition makes the fit unstable for the small size of
the \Bs sample, but can be used in the control channel \BdJpsiKst. The
differences found in the \Bd parameters with the two alternate
parameterizations are used as systematic uncertainties. The parameters
$\delta_{\parallel}$ fit to $\cos(\delta_{\parallel}) =
-0.960^{+0.021}_{-0.017}$ for the \Bd and to $\cos(\delta_{\parallel}) = -0.93
\pm 0.31$ (where the error corresponds to the positive one, being symmetrized)
for the \Bs. These parameters could in principle affect the efficiency
corrections, but it was found that the effect of different values of
$\delta_{\parallel}$ on the overall efficiency is negligible. A simulation
study of the fit pulls has shown that the errors on $f_L$ and $f_\parallel$ of
the \Bs decays are overestimated by a small amount ($\sim 10\%$) since they do
not follow exactly a Gaussian distribution, therefore the decision was taken to
quote an uncertainty which corresponds to an interval containing $68\%$ of the
generated experiments, rather than giving an error corresponding to a
log-likelihood interval of $0.5$. A slight bias observed in the pulls of
$f_\parallel$ in \Bs decays was accounted for by adding a systematic
uncertainty corresponding to 6\% of the statistical error.

The ratio of the two branching fractions is obtained from
\begin{equation}
\frac{\BRof\BsJpsiKst}{\BRof\BdJpsiKst} = \fdfs\,
\frac{\varepsilon^{\rm tot}_{\Bd}}{\varepsilon^{\rm tot}_{\Bs}}
\,\frac{\lambda_{\Bd}}{\lambda_{\Bs}} \,
\frac{f^{(d)}_{\Kstarz}}{f^{(s)}_{\Kstarz}} \,
\frac{N_{\Bs}}{N_{\Bd}},
\label{eq:br2}
\end{equation}
where $f_d$ $(f_s)$ is the probability of the $b$ quark to hadronize to $\Bd$
$(\Bs)$ mesons, $\varepsilon^{\rm tot}_{\Bd}/\varepsilon^{\rm tot}_{\Bs}$ is
the efficiency ratio, $\lambda_{\Bd}/\lambda_{\Bs}$ is the ratio of angular
corrections, $f^{(s)}_{\Kstarz}/f^{(d)}_{\Kstarz}$ is the ratio of \Kstarz
fractions and $N_{\Bs}/N_{\Bd}$ is the ratio of signal yields.  The value of
$f_d/f_s$ has been taken from Ref.~\cite{LHCb-PAPER-2011-018}. The efficiencies
in the ratio $\varepsilon_{\Bd}^{\rm tot}/\varepsilon_{\Bs}^{\rm tot}$ are
computed using simulation and receive two contributions: the efficiency of the
offline reconstruction (including geometrical acceptance) and selection cuts,
and the trigger efficiency on events that satisfy the analysis offline
selection criteria. The systematic uncertainty in the efficiency ratio is
negligible due to the similarity of the final states. Effects due to possible
differences in the decay time acceptance between data and simulation were found
to affect the efficiency ratio by less than 1 per mille. On the other hand,
since the efficiency depends on the angular distribution of the decay products,
correction factors $\lambda_{\Bd}$ and $\lambda_{\Bs}$ are applied to account
for the difference between the angular amplitudes used in simulation and those
measured in the data. The observed numbers of \Bd and \Bs decays, denoted by
$N_{\Bd}$ and $N_{\Bs}$, correspond to the number of \BsJpsiKpi and \BdJpsiKpi
decays with a $K\pi$ mass in a $\pm 40$ \mevcc window around the nominal \Kst
mass. This includes mostly the \Kst meson, but also an \swave component and the
interference between the \swave and \pwave components. The fraction of
candidates with a \Kstarz meson present is then
\begin{equation}
f_{\Kst} = \frac{\displaystyle\int_{\Omega}{\rm Acc}(\Omega) \;
\left. \frac{{\rm d}^3\Gamma}{{\rm d}\Omega} \right|_{|A_{\rm S}|=0} \;
{\rm d}\Omega} {\displaystyle\int_{\Omega}{\rm Acc}(\Omega) \;\;
\frac{{\rm d}^3\Gamma}{{\rm d}\Omega} \;\; {\rm d}\Omega},
\end{equation}
from which the ratio $f^{(s)}_{\Kstarz} / f^{(d)}_{\Kstarz} = 1.09 \pm 0.08$
follows. \tabref{tab:syst_br} summarizes all the numbers needed to compute the
ratio of branching fractions
\begin{equation} \nonumber
\frac{\BRof\BsJpsiKst}{\BRof\BdJpsiKst} =
\big( 3.43_{-0.36}^{+0.34} \pm 0.50 \big) \%.
\label{eq:branching_ratio}
\end{equation}

The contributions to the systematic uncertainty are also listed in
\tabref{tab:syst_br} and their relative magnitudes are: $1.2\%$ for the error
in the efficiency ratio; $2.5\%$ for the uncertainty on the transition point
($\alpha$) of the Crystal Ball function; $8.6\%$ for the parameterization of
the upper tail of the $\Bd$ peak; $3.9\%$ for the angular correction of the
efficiencies; $7.3\%$ for the uncertainty on the ratio
$f^{(s)}_{\Kstarz}/f^{(d)}_{\Kstarz}$ and $7.7\%$ for the uncertainty on
$f_d/f_s$. The errors are added in quadrature.
\begin{table}[t]
  \caption{\small Parameter values and errors for
  $\frac{\BRof\BsJpsiKst}{\BRof\BdJpsiKst}$.}
  \centering
  \setlength{\extrarowheight}{3pt}
  \begin{tabular}{c|c|c}
    Parameter & Name & Value  \\ 
    \hline
Hadronization fractions & $f_d / f_s$  & $3.75\pm0.29$ \\
Efficiency ratio & $\varepsilon^{\rm tot}_{\Bd}/\varepsilon^{\rm tot}_{\Bs}$ &
 $0.97 \pm 0.01$\\
Angular corrections  &$\lambda_{\Bd}/\lambda_{\Bs}$  & $1.01 \pm 0.04$ \\
Ratio of \Kstarz fractions & $f^{(s)}_{\Kstarz}/f^{(d)}_{\Kstarz}$  &
 $1.09 \pm 0.08$ \\
$B$ signal yields &$N_{\Bs}/N_{\Bd}$  & $\left( 8.5^{+0.9}_{-0.8} \pm 0.8
 \right)
\times 10^{-3}$ \\
  \end{tabular}
  \label{tab:syst_br}
\end{table}

Taking the value $\BRof\BdJpsiKst = (1.29 \pm 0.05 \pm 0.13) \times 10^{-3}$
from Ref.~\cite{Belle}
the following branching fraction is obtained,
\begin{equation} \nonumber
\BRof\BsJpsiKst = \big( 4.4_{-0.4}^{+0.5}\, \pm 0.8\, \big) \times 10^{-5}.
\end{equation}
This value is compatible with the CDF measurement~\cite{CDF} and is similar to
the naive quark spectator model prediction of Eq.~\eqref{eq:theory}, although
it is closer to the estimation in Ref.~\cite{faller}, $\BRof\BsJpsiKst \sim 2
\times \BRof\BdJpsiRho = (4.6\pm0.4)\times 10^{-5}$. The branching fraction
measured here is in fact the average of the \BsJpsiKst and $\Bsb \to \jpsi
\Kstarz$ branching fractions and corresponds to the time integrated quantity,
while theory predictions usually refer to the branching fraction at
$t=0$~\cite{deBruyn:2012wj}. In the case of \BsJpsiKst, the two differ by
$(\Delta\Gamma_s/2\Gamma_s)^2 = (0.77 \pm 0.25) \%$, where $\Delta\Gamma_s =
\Gamma_{\rm L} - \Gamma_{\rm H}$, $\Gamma_s = (\Gamma_{\rm L} + \Gamma_{\rm H})
/ 2$, and $\Gamma_{\rm L(H)}$ is the decay width of the light (heavy) \Bs-mass
eigenstate.

In conclusion, using $0.37$ fb$^{-1}$ of $pp$ collisions collected by the LHCb
detector at $\sqrt{s}$ = 7~TeV, a measurement of the \BsJpsiKst branching
fraction yields $\BF(\BsJpsiKst) = \big( 4.4_{-0.4}^{+0.5}\, \pm 0.8\, \big)
\times 10^{-5}$. In addition, an angular analysis of the decay products is
presented, which provides the first measurement of the \Kst polarization
fractions in this decay, giving $f_L = 0.50 \pm 0.08 \pm 0.02 $, $f_{\parallel}
= 0.19^{+0.10}_{-0.08} \pm 0.02$, and an \swave contribution of $|A_{\rm S}|^2
= 0.07^{+0.15}_{-0.07}$ in a $\pm 40\, \mevcc$ window around the \Kstarz mass.

\vspace{2mm}
We express our gratitude to our colleagues in the CERN accelerator departments
for the excellent performance of the LHC. We thank the technical and
administrative staff at CERN and at the LHCb institutes, and acknowledge
support from the National Agencies: CAPES, CNPq, FAPERJ and FINEP (Brazil);
CERN; NSFC (China); CNRS/IN2P3 (France); BMBF, DFG, HGF and MPG (Germany); SFI
(Ireland); INFN (Italy); FOM and NWO (The Netherlands); SCSR (Poland); ANCS
(Romania); MinES of Russia and Rosatom (Russia); MICINN, XuntaGal and GENCAT
(Spain); SNSF and SER (Switzerland); NAS Ukraine (Ukraine); STFC (United
Kingdom); NSF (USA). We also acknowledge the support received from the ERC
under FP7 and the Region Auvergne.




\ifx\mcitethebibliography\mciteundefinedmacro
\PackageError{LHCb.bst}{mciteplus.sty has not been loaded}
{This bibstyle requires the use of the mciteplus package.}\fi
\providecommand{\href}[2]{#2}


\end{document}